# CudaChain: A Practical GPU-accelerated 2D Convex Hull Algorithm

Gang Mei

Institute of Earth and Environmental Science, University of Freiburg

Albertstr.23B, D-79104, Freiburg im Breisgau, Germany

E-mail: gangmeiphd@gmail.com

**Abstract** This paper presents a practical GPU-accelerated convex hull algorithm and a novel *Sorting-based Preprocessing Approach* (SPA) for planar point sets. The proposed algorithm consists of two stages: (1) two rounds of preprocessing performed on the GPU and (2) the finalization of calculating the expected convex hull on the CPU. We first discard the interior points that locate inside a quadrilateral formed by four extreme points, and then distribute the remaining points into several (typically four) sub regions. For each subset of points, we first sort them in parallel, then perform the second round of discarding using SPA, and finally form a simple chain for the current remaining points. A simple polygon can be easily generated by directly connecting all the chains in sub regions. We at last obtain the expected convex hull of the input points by calculating the convex hull of the simple polygon. We use the library *Thrust* to realize the parallel sorting, reduction, and partitioning for better efficiency and simplicity. Experimental results show that our algorithm achieves 5x ~ 6x speedups over the Qhull implementation for 20M points. Thus, this algorithm is competitive in practical applications for its simplicity and satisfied efficiency.

## 1. Introduction

The finding of convex hulls is a fundamental issue in computer science, which has been extensively studied for many years. Several classic algorithms have been proposed, including the Graham scan [1], the Jarvis's march [2], the divide-and-conquer algorithm [3], the Andrew's monotone chain [4], the incremental approach [5], and the QuickHull [6].

Recently, to speed up the calculating of convex hulls for large sets of points, several efforts have been carried out to redesign and implement several commonly used CPU-based convex hull algorithms on the GPU. For example, Srikanth, et al. [7] and Srungarapu, et al. [8] parallelized the QuickHull algorithm [9] to accelerate the finding of two dimensional convex hulls. Based on the QuickHull approach, Stein, et al. [10] presented a novel parallel algorithm for computing the convex hull of a set of points in 3D using the CUDA programming model. Tang, et al. [11] developed a CPU-GPU hybrid algorithm to compute the convex hull of points in three or higher dimensional spaces.

Tzeng and Owens [12] presented a framework for accelerating the computing of convex hull in the divide-and-conquer fashion by taking advantage of QuickHull. Similarly, White and Wortman [13] described a pure GPU divide-and-conquer parallel algorithm for computing 3D convex hulls based on the Chan's minimalist 3D convex hull algorithm [14]. In [15], a novel algorithm is proposed to compute the convex hull of a point set in $R^3$ by exploiting the relationship between the Voronoi diagram and the convex hull. In addition, Gao, et al. [16] designed ffHull, a flip algorithm that allows nonrestrictive insertion of many vertices before any flipping of edges and maps well to the massively parallel nature of the modern GPU.

When calculating the convex hull of a set of points, an effective strategy for improving computational efficiency is to discard the interior points that have been exactly determined previously. This strategy is referred to as the *preprocessing/preconditioning* procedure. The most commonly used preprocessing approach is to form a convex polygon or polyhedron using several determined extreme points first and then discard those points that locate inside the convex polygon or polyhedron; see [10, 11]. The simplest case in two dimensions is to form a convex quadrilateral using four extreme points with min or max $x$ or $y$ coordinates and then check each point to determine whether it locates inside the quadrilateral; see [17]. Another quite recent effort for efficiently discarding interior points was introduced in [18].

In this paper, our objective is to design and implement an efficient and practical convex hull algorithm by exploiting the power of GPU. We make the following two contributions: (1) we propose a novel and effective *Sorting-based Preprocessing Approach* (SPA) for discarding interior points; (2) we present an efficient GPU-accelerated algorithm for finding the convex hulls of planar point sets based on the algorithm SPA.

The proposed convex hull algorithm consists of two stages: (1) two rounds of preprocessing performed on the GPU and (2) the finalization of calculating the expected convex hull on the CPU. We first discard the interior points that locate inside a quadrilateral formed by four extreme points, and then distribute the remaining points into several (typically four) sub regions. For each subset of points, we first sort them in parallel, then perform the second round of discarding using SPA, and finally form a simple chain for the current remaining points. A simple polygon can be easily generated by directly connecting all the chains in sub regions. We at last obtain the expected convex hull of the input points by calculating the convex hull of the simple polygon using the Melkman algorithm [19].

Our algorithm is implemented by heavily taking advantage of the library *Thrust* [20] for better efficiency and simplicity. In our implementation, we directly use the very efficient data parallel primitives such as parallel sorting, reduction, and partitioning that are provided by Thrust. The use of the library Thrust makes our implementation simple to develop.

We test our algorithm against the Qhull library [21] on various datasets of different sizes using two machines. Experimental results show that our algorithm achieves 5x ~ 6x speedups over the Qhull implementation for 20M points. Thus, this algorithm is competitive in practical applications for its simplicity and satisfied efficiency performance.

## 2. Methods

### 2.1 Algorithm Design

The proposed GPU-accelerated algorithm is designed on the basis of the fast convex hull algorithm introduced by Akl and Toussaint [17]. The procedure of our algorithm mainly consists of three steps: (1) we first carry out a first round of preprocessing by discarding those points locating inside a quadrilateral formed by four extreme points. This commonly used strategy of preprocessing was described in [17]; (2) we then distribute the remaining points into several (typically four) sub regions, sort the points in the same region according to their coordinates, and perform a novel *Sorting-based Preprocessing Approach* (SPA) to in further discard interior points for each sub region and form a simple polygon; (3) we finally employ Melkman's algorithm [19] to calculate the convex hull of the simple polygon. The obtained convex hull is exactly the expected convex hull of the input point set. The first and second steps of our algorithm are performed on the GPU, while the third is carried out on the CPU.

More specifically, the procedure of the proposed algorithm is listed as follows:

1) Find four extreme points that have the max or min *x* and *y* coordinates by parallel reduction, denote them as $P_{minx}$, $P_{maxx}$, $P_{miny}$, $P_{maxy}$
2) Determine the distribution of all points in parallel, and discard the points locating inside the quadrilateral formed by $P_{minx}$, $P_{miny}$, $P_{maxx}$, and $P_{maxy}$
3) Denote the subset of points locating in the four sub regions, i.e., the lower left, lower right, upper right, and upper left as $S_{R1}$, $S_{R2}$, $S_{R3}$, and $S_{R4}$, respectively
4) Sort $S_{R1}$, $S_{R2}$, $S_{R3}$, and $S_{R4}$ separately in parallel; see Table 1 for the orders of sorting
5) Perform the SPA for $S_{R1}$, $S_{R2}$, $S_{R3}$, and $S_{R4}$ to discard interior points in further, and form a simple chain for the remaining points in each sub region
6) Form a simple polygon by connecting those four chains in counterclockwise (CCW)
7) Find the convex hull of the simple polygon using Melkman's algorithm [19]

In the above procedure, the most commonly used strategy for discarding interior points is first carried out (i.e., the step 1 and step 2); and then those remaining points are divided into four subsets. After that, each subset of points is sorted separately. The key step in this procedure is the second round of discarding interior points and the forming of a simple chain for each subset. A simple polygon can be easily created by directly connecting the chains; and the expected convex hull can be found using Melkman's algorithm [19] which is specifically designed for calculating the convex hull of a simple polygon.

**Table 1** Regions and corresponding rules for discarding interior points

| Region | First point | Last point | Order of *x* coordinates | Order of *y* coordinates | Rule for discarding |
|---|---|---|---|---|---|
| Lower left (R$_1$) | $P_{minx}$ | $P_{miny}$ | Ascending | Descending | Sort *x* in ascending order first Discard point *P* according to the rule listed in Figure 2 (a) |
| Lower right (R$_2$) | $P_{miny}$ | $P_{maxx}$ | Ascending | Ascending | Sort *y* in ascending order first Discard point *P* according to the rule listed in Figure 2 (b) |
| Upper right (R$_3$) | $P_{maxx}$ | $P_{maxy}$ | Descending | Ascending | Sort *x* in descending order Discard point *P* according to the rule listed in Figure 2 (c) |
| Upper left (R$_4$) | $P_{maxy}$ | $P_{minx}$ | Descending | Descending | Sort *y* in descending order Discard point *P* according to the rule listed in Figure 2 (d) |

**2.1.1 Step 1 : Points' Distribution and the First Round of Discarding**

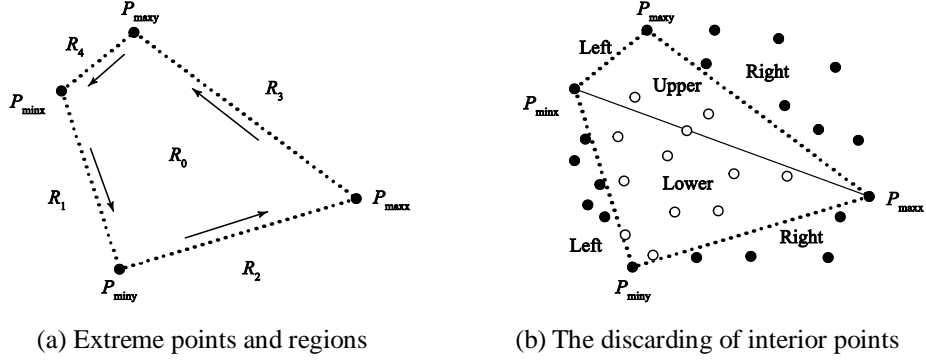

(a) Extreme points and regions      (b) The discarding of interior points

**Figure 1** The distribution of points and the first round of discarding

The strategy of discarding the interior points locating inside a quadrilateral formed by four extreme points is straightforward; see Figure 1. There is no need to describe this strategy in more details. The only remarkable issue is that: to reduce the computational cost, in the process of checking whether a point is interior (i.e., locating in the region $R_0$ in Figure 1(a)), the distribution of those non-interior points can also be easily determined. For all points, we use the following simple method to determine their distributions:

(1) if point $P$ lies on the left side of the directed line $P_{minx}P_{miny}$, then it falls in the region $R_1$;

(2) else if $P$ lies on the left side of the directed line $P_{miny}P_{maxx}$, then it falls in the region $R_2$;

(3) else if $P$ lies on the left side of the directed line $P_{maxx}P_{maxy}$, then it falls in the region $R_3$;

(4) else if $P$ lies on the left side of the directed line $P_{maxy}P_{minx}$, then it falls in the region $R_4$;

(5) else $P$ falls in the region $R_0$

After the above procedure of determination, all points are distributed into five regions. Those points in the region $R_0$ are interior ones, and need to be directly discarded in the step, while the remaining points in the other four regions should be taken into consideration for calculating the convex hull.

### 2.1.2 Step 2 : Second Round of Discarding and Forming Simple Polygon

**(a)**
$n_1$ = number of points in the region $R_1$
$t = y_0$
**for** $i = 1$ **to** $i < n_1 - 1$ **do**
   **if** $y_i > t$, **then** point $P_i$ is interior
   **else** $t = y_i$
**end**

**(b)**
$n_2$ = number of points in the region $R_2$
$t = x_0$
**for** $i = 1$ **to** $i < n_2 - 1$ **do**
   **if** $x_i < t$, **then** point $P_i$ is interior
   **else** $t = x_i$
**end**

**(c)**
$n_3$ = number of points in the region $R_3$
$t = y_0$
**for** $i = 1$ **to** $i < n_3 - 1$ **do**
   **if** $y_i < t$, **then** point $P_i$ is interior
   **else** $t = y_i$
**end**

**(d)**
$n_4$ = number of points in the region $R_4$
$t = x_0$
**for** $i = 1$ **to** $i < n_4 - 1$ **do**
   **if** $x_i > t$, **then** point $P_i$ is interior
   **else** $t = x_i$
**end**

**Figure 2** The rules for discarding interior points of the SPA

This section will describe a novel sorting-based preprocessing approach that is specifically applicable to the previously sorted points. We term this method as the SPA. The rules for discarding interior points in those four sub regions, i.e., lower left ($R_1$), lower right ($R_2$), upper right ($R_3$), and upper left ($R_4$), are presented in Figure 2.

The correctness of SPA for each sub region is demonstrated in Figure 3. For the region $R_1$, the first point and the last point are the $P_{minx}$ and $P_{miny}$, respectively; see Figure 3(a). Assuming the point $P_i$ has been determined to be non-interior, and now we check the point $P$ according to the relationship between $P_i$ and $P$. Since that all the points in the region $R_1$ have been sorted in the ascending order of $x$ coordinates, thus $x_P > x_{Pi}$; and if $y_{Pi}$ is larger than $y_P$, then the point $P$ must be located in the shaded triangular area; and obviously it also falls in the triangle formed by three points $P_i$, $P_{miny}$, and $P_{minx}$. Hence, the point $P$ must be an interior one and needs to be discarded. The correctness of SPA for other three regions can also be explained similarly.

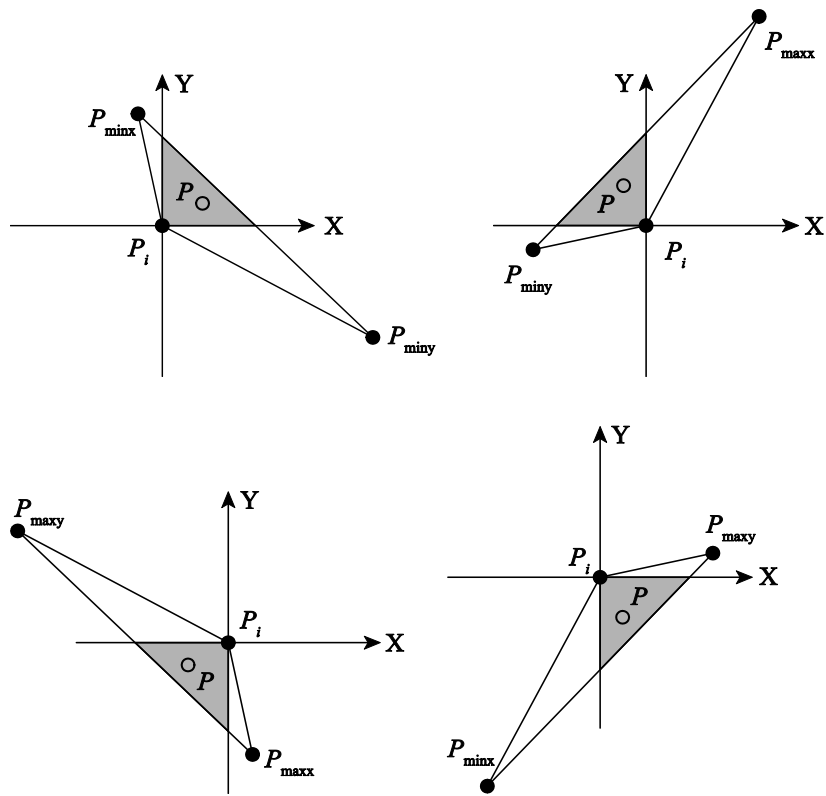

**Figure 3** Demonstrations for the correctness of discarding interior points in the method SPA. (a): the region $R_1$; (b) the region $R_2$; (c) the region $R_3$; (d) the region $R_4$

We also take the forming of the chain in the upper left region ($R_3$) for an example; see Figure 4. Previously, seven points have been sorted in the descending order of $x$. We first check the point $P_1$; and obviously it is not an interior point according the rule presented in the Figure 2 (c). Similarly, we also find that the point $P_2$ is an exterior point and needs to be kept. However, the point $P_3$ is an interior point since its $y$ coordinate is less than that of the point $P_2$; and obviously, the point $P_3$ locates inside the triangle formed by the first point (i.e., the point $P_{maxx}$), the last point (i.e., the point $P_{maxy}$), and $P_2$. After discarding the point $P_3$, the point $P_4$ can also be discarded, while both the points $P_5$ and $P_6$ are not exterior points. However, since that the coordinate $y$ of the point $P_7$ is less than that of the point $P_6$, it also should be removed.

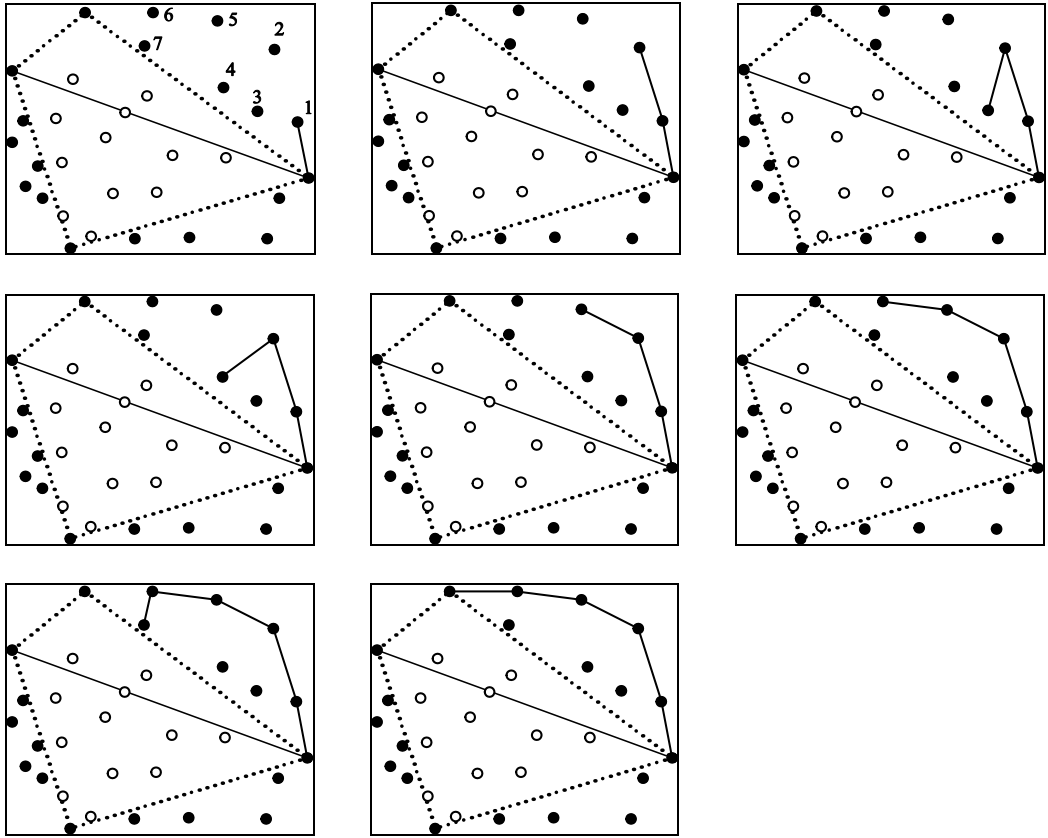

**Figure 4** A simple example of forming the chain in the upper left region

**2.1.3 Step 3: Calculating the Convex Hull of Simple Polygon**

The output of the previous step is a simple polygon, which is also an approximate convex hull. To calculate the exact convex hull of the input point set, we select the fast algorithm introduced by Melkman [19] to compute the convex hull of the simple polygon. The convex hull of the simple polygon is the convex hull of the input data set; see Figure 5.

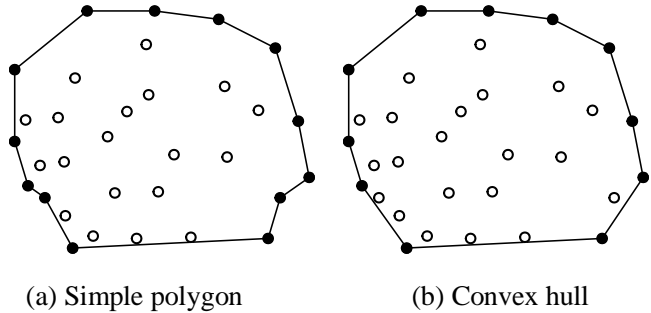

(a) Simple polygon  (b) Convex hull
**Figure 5** The convex hull of a simple polygon

## 2.2 Implementation Details

In this section, we describe more details about the implementation of our algorithm. The implementation has both the CPU side (host) and the GPU side (device) code. The code on the CPU side is developed to compute the convex hull of a simple polygon, which is relatively simple and easy to implement when compared to the code on the GPU side. Thus, we focus on the development of the GPU side code.

The implementation on the GPU side is developed by heavily taking advantage of the library *Thrust* for better efficiency and simplicity when using the data-parallel algorithm primitives. Currently, several GPU-accelerated libraries have been designed to provide data-parallel algorithm primitives such as parallel scan, parallel sort and parallel reduction. Such libraries include Thrust [20], CUDPP[22], and CUB[23]. Since that the library Thrust has been integrated in CUDA since version 4.0, it is very easy and convenient to use Thrust directly in CUDA. Hence, we select the library Thrust rather than the other two libraries.

```
1   // input : thrust::host_vector<float> x, y
2   // output: thrust::host_vector<float> hull_x, hull_y
3
4   int n = x.size();   // Number of input points
5
6   thrust::device_vector<float> d_x = x;    // Array for storing x on GPU
7   thrust::device_vector<float> d_y = y;    // Array for storing y on GPU
8   thrust::device_vector<int>   d_pos(n);   // Indicators of distribution
9
10  typedef thrust::device_vector<float>::iterator  floatIter;
11  floatIter minx = thrust::min_element(d_x.begin(), d_x.end());  // min X
12  floatIter maxx = thrust::max_element(d_x.begin(), d_x.end());  // max X
13  floatIter miny = thrust::min_element(d_y.begin(), d_y.end());  // min Y
14  floatIter maxy = thrust::max_element(d_y.begin(), d_y.end());  // max Y
15
16  thrust::device_vector<float> d_extreme_x(4);  // X of extreme points
17  thrust::device_vector<float> d_extreme_y(4);  // Y of extreme points
18
19  d_extreme_x[0] = *minx;  d_extreme_y[0] = d_y[minx - d_x.begin()];
20  d_extreme_x[1] = d_x[miny - d_y.begin()];  d_extreme_y[1] = *miny;
21  d_extreme_x[2] = *maxx;  d_extreme_y[2] = d_y[maxx - d_x.begin()];
22  d_extreme_x[3] = d_x[maxy - d_y.begin()];  d_extreme_y[3] = *maxy;
23
24  float * d_extreme_x_ptr = thrust::raw_pointer_cast(&d_extreme_x[0]);
25  float * d_extreme_y_ptr = thrust::raw_pointer_cast(&d_extreme_y[0]);
26  float * d_x_ptr = thrust::raw_pointer_cast(&d_x[0]);
27  float * d_y_ptr = thrust::raw_pointer_cast(&d_y[0]);
28  int * d_pos_ptr = thrust::raw_pointer_cast(&d_pos[0]);
29
30  // Kernel for determining points' distribution
31  kernelPreprocess<<<(n + 1023) / 1024, 1024>>>(d_extreme_x_ptr, d_extreme_y_ptr,
32                                               d_x_ptr, d_y_ptr, d_pos_ptr, n);
33
34  // Create some zip_iterators
35  typedef thrust::device_vector<int>::iterator          intIter;
36  typedef thrust::tuple<floatIter, floatIter, intIter>  pointIterTuple;
37  typedef thrust::zip_iterator<pointIterTuple>          pointIter;
38  pointIter P_first = thrust::make_zip_iterator(
39                         make_tuple(d_x.begin(), d_y.begin(), d_pos.begin()));
40  pointIter P_last  = thrust::make_zip_iterator(
41                         make_tuple(d_x.end(),   d_y.end(),   d_pos.end()));
42
43  // Partition to gather points in the same region together
44  pointIter first_of_R0 = thrust::partition(P_first, P_last, is_interior);
45  pointIter first_of_R2 = thrust::partition(P_first, first_of_R0-1, is_region_1);
46  pointIter first_of_R3 = thrust::partition(first_of_R2, first_of_R0-1, is_region_2);
47  pointIter first_of_R4 = thrust::partition(first_of_R3, first_of_R0-1, is_region_3);
48
49  // Sort Partly in each region using sort_by_key()
50  // Region 1 : ascending X      Region 3 : descending X
51  // Region 2 : ascending Y      Region 4 : descending Y
52     thrust::sort_by_key(...); ...
53
54  // Kernels for 2nd round of discarding (SPA)
55  kernelCheck_R1<<<1, BLOCK_SIZE>>>(d_y_ptr, d_pos_ptr, ...);  // Only Y
56  kernelCheck_R2<<<1, BLOCK_SIZE>>>(d_x_ptr, d_pos_ptr, ...);  // Only X
57  kernelCheck_R3<<<1, BLOCK_SIZE>>>(d_y_ptr, d_pos_ptr, ...);  // Only Y
58  kernelCheck_R4<<<1, BLOCK_SIZE>>>(d_x_ptr, d_pos_ptr, ...);  // Only X
59
60  // Partition again and then Copy
61  pointIter P_valid = thrust::stable_partition(P_first, first_of_R0, is_interior());
62  n = P_valid - P_first;   // Number of vertices of the output Simple Polygon
63  thrust::copy_n(thrust::get<0>(pos_R1), n, hull_x.begin());
64  thrust::copy_n(thrust::get<1>(pos_R1), n, hull_y.begin());
```

**Figure 6** The implementation of the proposed algorithm (CudaChain)

### 2.2.1 Performing the First Round of Discarding on the GPU

The first step of discarding the interior points locating inside the quadrilateral formed by four extreme points is to find those points with the min or max $x / y$ coordinates. In sequential programming pattern, a loop over all input points needs to be carried out to find the min or max values. In parallel programming pattern, the finding of min or max values in a vector can be efficiently achieved by performing a parallel reduction. Thrust provides such common data-parallel primitive and several easy-to-use interface functions. We use two functions, i.e., `thrust::min_element()` and `thrust::max_element()`, to efficiently find the min and max coordinates of all points in parallel; see lines 11 - 14 in Figure 6.

To avoid the transformation between device memory and host memory, we directly obtain the memory addresses of the coordinates of extreme points and all input points that resides on the GPU using the function `thrust::raw_pointer_cast()`, and then pass them as the launch arguments for the kernel `kernelPreprocess`; see lines 24 - 28 in Figure 6.

We design a kernel, `kernelPreprocess`, to determine in which region a point falls. Each thread is responsible for calculating the position of a point; and the results are stored in an array `d_pos[n]`. The method for determining the distribution of points is introduced in Section 2.1.1. We use an integer as an indicator of the position. For example, if a point $P_i$ locates inside the region $R_1$, then the value `d_pos[i]` is set to 1; and the indicator value of an interior point is 0. All the points that are not in the region $R_0$ are called exterior or remaining points. A simple example is presented in Figure 7(a).

### 2.2.2 Performing the Second Round of Discarding on the GPU

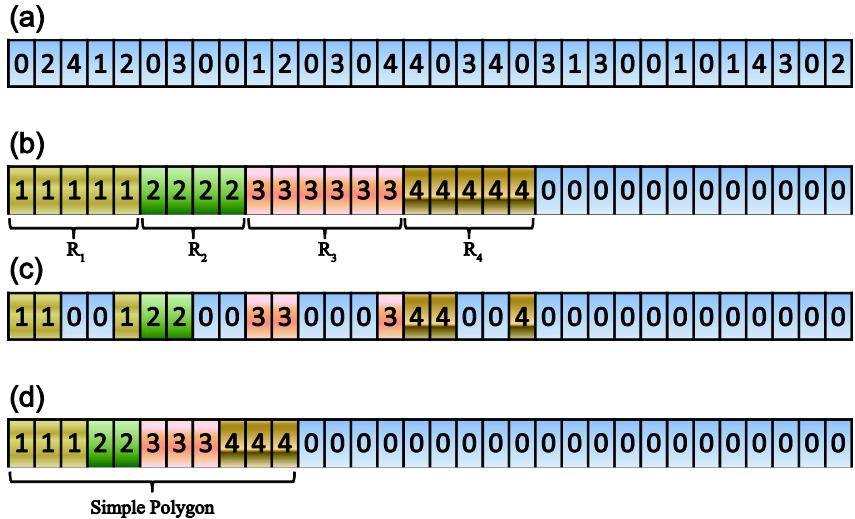

**Figure 7** The second round of discarding interior points

To carry out the second round of discarding interior points, we: (1) perform four parallel partitioning for all points according to their positions, (2) sort the points in each region separately, (3) invoke a kernel for each region to discard the interior points using the method SPA, and (4) perform another parallel partitioning for all exterior points.

The first step, parallel partitioning, is carried out to gather those points in the same region together for subsequent procedure of sorting. After partitioning, the points that locate in the same region reside in a consecutive segment; see Figure 7(b). In this case, parallel sorting can be performed for each segment of points. Noticeably, we choose to partition and sort each segment of points in place to minimize the cost of memory space.

After parallel sorting, we design a kernel for each region to discard the interior points using the method SPA. There is only one thread block within the kernel's thread grid. Each thread in the only thread block is responsible for checking consecutive ($m$ + BLOCK_SIZE - 1) / BLOCK_SIZE points in the same region, where $m$ is the number of points in a region for being checked, and BLOCK_SIZE represents the number of threads in the only block. In our implementation, we set BLOCK_SZIE to 1024 according to the compute capability of the adopted GPU. After checking and discarding interior points using SPA, some previous exterior points have been determined as interior ones; and their corresponding indicator values are modified to 0; see Figure 7(c).

In our implementation, we only allocate one thread block in the discarding of interior points using the proposed method SPA due to the data dependency issue in the discarding. When checking whether a point in a specific region such as the region $R_1$, the $y$ coordinate of the point being checked is compared to that of the current last point of the formed chain; see Figure 2(a). This means the checking for a point, e.g., $P_i$, depends on the checking of the previous point $P_{i-1}$. It also means the checking for a set of consecutive points can only be performed in a sequential pattern. However, it is able to first divide a large set of consecutive points into some smaller subsets of consecutive points, and then perform the checking in parallel for each subset of points separately. This solution is in the divide-and-conquer fashion. We adopt this solution; but we cannot determine the optimal size of a subset of points or the number of all subsets. Thus, we decide to divide a large set of consecutive points into BLOCK_SIZE subsets, while each subset contains ($m$ + BLOCK_SIZE - 1) / BLOCK_SIZE points; and then, we carry out the checking for all the BLOCK_SIZE subsets in parallel.

The final step is to re-partition and copy the coordinates of the exterior points in current stage for outputting. Noticeably, to preserve the relative order of the sorted points, we use the function `thrust::stable_partition()` rather than `thrust::partition()` to compact the exterior points; see lines 60 - 64 in Figure 6. After the stable partitioning, the remaining exterior points are stored consecutively and can be easily copied in parallel for being used on the host side (on the CPU); see Figure 7(d).

## 3. Results

We have tested our algorithm against the Qhull library [21] on various datasets of different sizes using two machines. The first machine features an Intel i7-3610QM processor (2.30GHz), 6GB of memory and a NVIDIA GeForce GTX660M graphics card. The other machine has an Intel i5-3470 processor (3.20GHz), 8GB of memory and a NVIDIA GeForce GT640 (GDDR5) graphics card. The graphics card GTX 660M has 2GB of RAM and 384 cores; and the GT640 graphics card has 1GB of RAM and 384 cores. We have used the CUDA toolkit version 5.5 on Window 7 Professional for evaluating all the experimental tests.

We have created two groups of datasets for testing. The first group includes 8 sets of randomly distributed points in a square that are generated using the `rbox` component in Qhull. The second group consists of 10 point sets that are derived from 3D mesh models by projecting the vertices of each 3D model onto the XY plane. These mesh models presented in Figure 8 are directly obtained from the Stanford 3D Scanning Repository[1] and the GIT Large Geometry Models Archive[2].

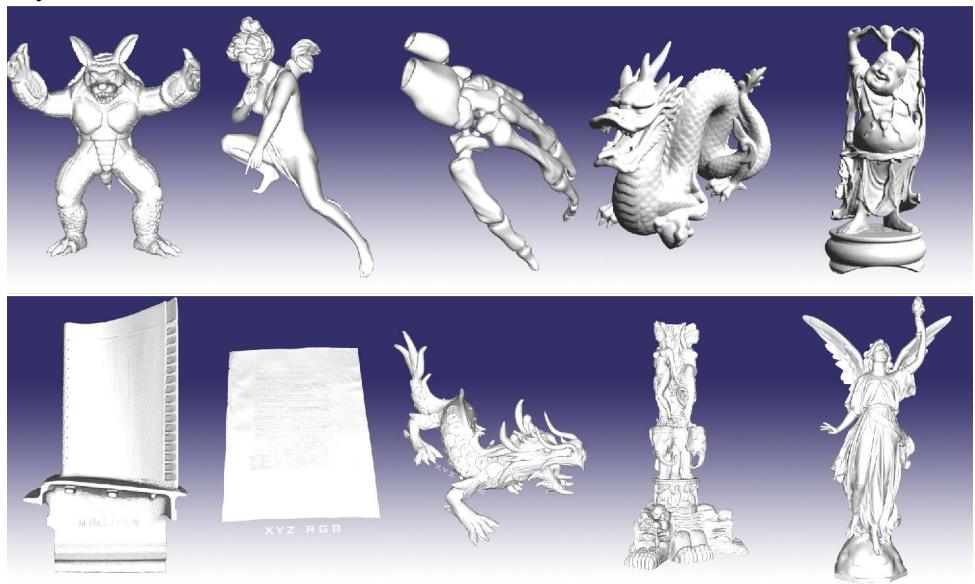

**Figure 8** 3D mesh models from Stanford 3D Scanning Repository and GIT Large Geometry Models Archive. From the left to the right, the models are: Armadillo, Angel, Skeleton Hand, Dragon, Happy Buddha, Turbine Blade, Vellum Manuscript, Asian Dragon, Thai Statue, and Lucy

### 3.1 Efficiency on the GTX 660M

The running time on the GPU GTX 660.M of two groups of testing data, i.e., the group of randomly distributed point sets and the group of point sets derived from 3D models, is listed in Table 2 and Table 3, respectively. To evaluate the computation load between the GPU side and the CPU side of our algorithm, we count the running time separately for both of the two sides and calculate the workload percentage of the CPU side.

For both the two groups of experimental tests, the experimental results show that: for small size of testing data, the Qhull is faster than the proposed CudaChain, while CudaChain is much faster than Qhull for the large size of testing data. The speedups of CudaChain over Qhull become larger with the increasing of the data size. The speedup is about 3x~4x on average and 5x~6x in the best cases.

The workload percentage of the CPU side is much smaller than that on the GPU side; and it decreases for the group of randomly point sets when the data size increases. In addition, the workload percentage of the CPU side is usually less that 10%, except for the test of the model Happy Buddha.

**Table 2** Comparison of running time (/ms) for randomly distributed point sets on GTX 660M

---
[1] http://www-graphics.stanford.edu/data/3Dscanrep/

[2] http://www.cc.gatech.edu/projects/large_models/

| Size | Qhull | CudaChain | | | | Speedup |
|---|---|---|---|---|---|---|
| | | Total | GPU | CPU | CPU(%) | |
| 100k | 27 | 42.5 | 39.6 | 2.9 | 6.82 | 0.64 |
| 200k | 52 | 45.9 | 43.1 | 2.8 | 6.10 | 1.13 |
| 500k | 124 | 65.6 | 61.4 | 4.2 | 6.40 | 1.89 |
| 1M | 237 | 75.0 | 70.8 | 4.2 | 5.60 | 3.16 |
| 2M | 426 | 129.1 | 123.2 | 5.9 | 4.57 | 3.30 |
| 5M | 605 | 174.8 | 169.4 | 5.4 | 3.09 | 3.46 |
| 10M | 1171 | 351.8 | 345.9 | 5.9 | 1.68 | 3.33 |
| 20M | 2353 | 587.4 | 581.9 | 5.5 | 0.94 | 4.01 |

**Table 3** Comparison of running time (/ms) for point sets derived from 3D models on GTX 660M

| 3D Model | Size | Qhull | CudaChain | | | | Speedup |
|---|---|---|---|---|---|---|---|
| | | | Total | GPU | CPU | CPU(%) | |
| Armadillo | 172k | 47 | 39.7 | 37.6 | 2.1 | 5.29 | 1.2 |
| Angel | 237k | 51 | 41.6 | 38.7 | 2.9 | 6.97 | 1.2 |
| Skeleton Hand | 327k | 77 | 45.4 | 41.5 | 3.9 | 8.59 | 1.7 |
| Dragon | 437k | 98 | 59.8 | 53.9 | 5.9 | 9.87 | 1.6 |
| Happy Buddha | 543k | 123 | 68.7 | 59.6 | 9.1 | 13.25 | 1.8 |
| Turbine Blade | 882k | 202 | 73.9 | 67.5 | 6.4 | 8.66 | 2.7 |
| Vellum Manuscript | 2M | 392 | 90.6 | 86.9 | 3.7 | 4.08 | 4.3 |
| Asian Dragon | 3M | 492 | 101.7 | 97.9 | 3.8 | 3.74 | 4.8 |
| Thai Statue | 5M | 547 | 106.0 | 102.4 | 3.6 | 3.40 | 5.2 |
| Lucy | 14M | 1481 | 245.2 | 240.9 | 4.3 | 1.75 | 6.0 |

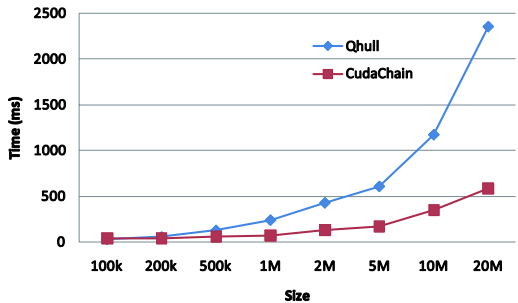
(a) Randomly distributed points

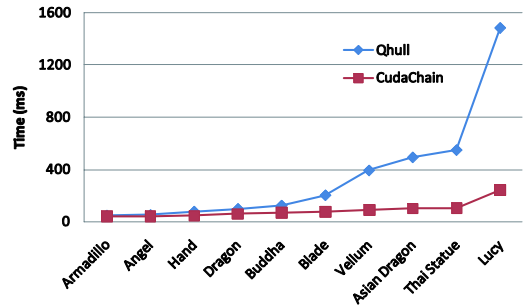
(b) Points derived from 3D models

**Figure 9** The efficiency of CudaChain on the GPU GTX 660M against CPU-based Qhull using the same datasets and the same machine

## 3.2 Efficiency on the GT 640

On the machine with the GPU GT640, the running time of two groups of testing data is listed in Table 4 and Table 5. Similar to those experimental results obtained on the machine with the GTX 660M, for small size of testing data, the Qhull is also faster than our algorithm CudaChain, while CudaChain is much faster than Qhull for the large size of testing data. The speedups of CudaChain over Qhull also become larger with the increasing of the data sizes.

The speedup is about 3x~4x on average and 4x~5x in the best cases; however, for the largest model Lucy, the speedup is 5.2x on the GT 640, while it is 6x on the GTX 660M.

The experimental results obtained on the machine with the GT 640 also indicate that: the workload percentage of the CPU side is much smaller than that of the GPU side; and it decreases for the group of randomly point sets when the data size increases. The behaviors are the same as those on the GTX 660M. Furthermore, the workload percentage of the CPU side is usually also less that 10%, except for the test of the model Happy Buddha.

**Table 4** Comparison of running time (/ms) for randomly distributed point sets on GT 640

| Size | Qhull | CudaChain | | | | Speedup |
|---|---|---|---|---|---|---|
| | | Total | GPU | CPU | CPU(%) | |
| 100k | 15 | 25.5 | 24.2 | 1.3 | 5.10 | 0.59 |
| 200k | 16 | 29.1 | 27.9 | 1.2 | 4.12 | 0.55 |
| 500k | 47 | 40.4 | 38.4 | 2.0 | 4.95 | 1.16 |
| 1M | 109 | 46.9 | 44.6 | 2.3 | 4.90 | 2.32 |
| 2M | 202 | 83.5 | 81.2 | 2.3 | 2.75 | 2.42 |
| 5M | 515 | 147.0 | 144.5 | 2.5 | 1.70 | 3.50 |
| 10M | 1034 | 321.9 | 319.7 | 2.2 | 0.68 | 3.21 |
| 20M | 2215 | 544.4 | 541.5 | 2.9 | 0.53 | 4.07 |

**Table 5** Comparison of running time (/ms) for point sets derived from 3D models on GT 640

| 3D Model | Size | Qhull | CudaChain | | | | Speedup |
|---|---|---|---|---|---|---|---|
| | | | Total | GPU | CPU | CPU(%) | |
| Armadillo | 172k | 15 | 26.5 | 25.0 | 1.5 | 5.66 | 0.6 |
| Angel | 237k | 16 | 28.0 | 26.4 | 1.6 | 5.71 | 0.6 |
| Skeleton Hand | 327k | 31 | 29.6 | 27.8 | 1.8 | 6.08 | 1.0 |
| Dragon | 437k | 47 | 35.1 | 31.8 | 3.3 | 9.40 | 1.3 |
| Happy Buddha | 543k | 62 | 42.1 | 37.4 | 4.7 | 11.16 | 1.5 |
| Turbine Blade | 882k | 78 | 46.3 | 42.8 | 3.5 | 7.56 | 1.7 |
| Vellum Manuscript | 2M | 218 | 78.5 | 75.4 | 3.1 | 3.95 | 2.8 |
| Asian Dragon | 3M | 343 | 102.5 | 98.8 | 3.7 | 3.61 | 3.3 |
| Thai Statue | 5M | 468 | 105.5 | 101.9 | 3.6 | 3.41 | 4.4 |
| Lucy | 14M | 1295 | 248.8 | 244.5 | 4.3 | 1.73 | 5.2 |

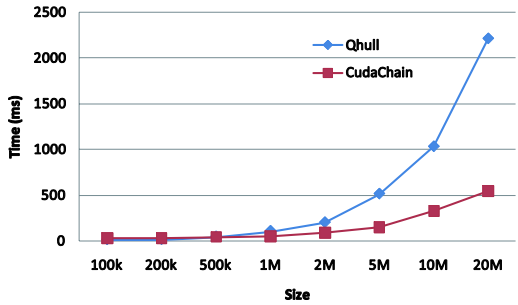
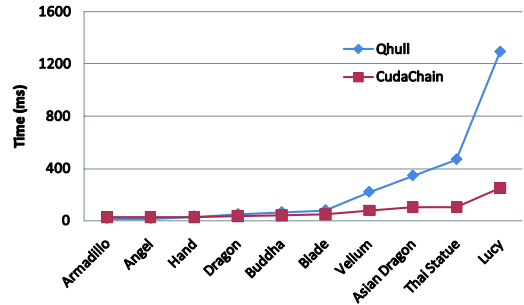

(a) Randomly distributed points        (b) Points derived from 3D models

**Figure 10** The efficiency of CudaChain on the GPU GT 640 against CPU-based Qhull using the same datasets and the same machine

## 3.3 Effectiveness of Discarding Interior Points

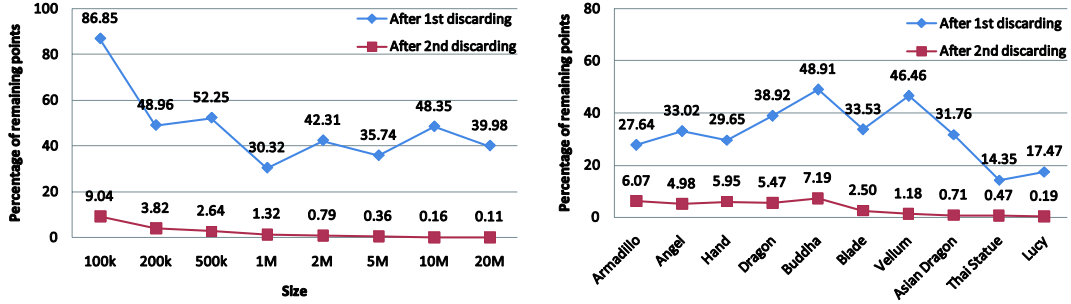

(a) Randomly distributed points  (b) Points derived from 3D models

**Figure 11** The effectiveness of two rounds of discarding interior points. (First round of discarding: removing the points inside the quadrilateral formed by four extreme points. Second round of discarding: removing interior points according to the rules of the method SPA)

There are two rounds of discarding in our algorithm. To evaluate the effectiveness of the proposed preprocessing method SPA, we count the remaining points after each round of discarding, and compare the effectiveness of two rounds of discarding. The results presented in Figure 11 show that SPA can dramatically reduce the number of remaining points and thus improve the overall efficiency of CudaChain. In addition, the effectiveness of discarding interior points by SPA becomes better with the increasing of the data size.

## 4. Discussion

### 4.1 Comparison

We have tested our algorithm CudaChain on two different machines with different GPUs. The efficiency performances of our algorithm on the two machines are almost the same. This result is due to the fact that the two GPUs, i.e., GTX 660M and GT 640, have the similar compute capability. However, the speedups of our algorithm over the implementation Qhull on two machines slightly differ. This behavior is lead by the different efficiency performance of CPU-based Qhull on the two machines.

Compared to other GPU-accelerated convex hull algorithms such as those implemented by parallelizing the QuickHull algorithm on the GPU [7, 8, 10, 12], the main advantage of our algorithm is that it is very easy to implement, which is mainly due to (1) the use of the library Thrust and (2) relatively less data dependencies. The data-parallel primitives such as parallel sorting and parallel reduction provided by Thrust are very efficient and easy to use; we can directly use these primitives in CUDA to realize our implementation without too many efforts. In addition, in our algorithm the only step that has data dependency is the checking and discarding interior points using SPA. Other steps or procedures can be very well mapped to the massively parallel nature of the modern GPU. This feature of having less data dependencies also makes our algorithm simple and easy to implement in practical applications.

## 4.2 Complexity and Correctness

The time complexity in the worst case of the second round of discarding is $O(n\log n)$ due to the sorting of points. Both the first round of discarding and the calculating of convex hull of a simple polygon run in $O(n)$. Thus, the worst case time complexity of the entire algorithm is $O(n\log n)$.

The space requirement of our algorithm is also efficient. Only three arrays for storing all the input points' coordinates and positions need to be allocated on the GPU. The parallel sorting, parallel reduction, and parallel partitioning completely operate on those three arrays in place without needing to explicitly allocate any additional global memory. In addition, to avoid the transformation from device memory and host memory and then back to device memory when invoking user-designed kernels, we directly obtain the memory addresses of those three arrays that resides on the GPU using the function `thrust::raw_pointer_cast()`, and then pass the addresses as the launch arguments for kernels.

The correctness of our algorithm can obviously be guaranteed. It is clear that: (1) in the calculating of convex hulls, any potential extreme points should not be discarded; and (2) any points that have been identified as interior ones can be discarded. As mentioned several times, there are two rounds of discarding in our algorithm. In the first round of discarding, those points locating in the quadrilateral formed by extreme points are definitely the interior ones and can be discarded. In the second round of discarding, we have proved that the points detected as the interior using the proposed preprocessing method SPA is reasonable; see Figure 3. Thus, this discarding can also be guaranteed to be correct. After two rounds of discarding, all the remaining points are used to calculate the expected convex hull.

In short, our algorithm can be guaranteed to be correct since we (1) only remove recognized interior points and (2) preserve all remaining points to avoid discarding any potential extreme points.

## 4.3 Limitation

The first shortcoming of our algorithm is that: the efficiency of discarding interior points using SPA within a single thread block cannot be guaranteed to be the highest. Probably, to hide memory latency and improve the efficiency, it needs to allocate several thread blocks in the discarding of interior points using the method SPA; and each thread is responsible for checking and discarding interior points for a small subset of consecutive points. However, the optical number of consecutive points that assigned to be checked in each thread to generate the highest efficiency cannot be determined; this is due to (1) the different data size of input points and the number of remaining points after the first round of discarding, and (2) the compute capability of the adopted GPU. Thus, it needs more experimental tests to determine the optical number of points that are assigned to each thread.

Another limitation is that: when all the input points are initially extreme points, e.g., when all points exactly locate on a circle, two rounds of discarding interior points will be wasteful since there are no interior points that will be found and removed. All the input points will be kept and used to calculate the convex hull on the CPU using the Melkman algorithm [19]. Hence, the overall execution in this case could be very slow.

## 5. Conclusion

We have proposed a novel sorting-based preprocessing approach (SPA) and an efficient algorithm for calculating the convex hulls of planar point sets on the GPU. Our algorithm consists of two rounds of preprocessing procedures performed on the GPU and the finalization of calculating the expected convex hull on the CPU. We have used the library Thrust to realize the parallel sorting, reduction, and partitioning for better efficiency and simplicity. We have tested our algorithm against the Qhull library on various datasets of different sizes using two machines. Experimental results show that our algorithm achieves the speedups of 3x~4x on average and 5x~6x in the best cases. We believe that our algorithm is competitive in practical applications for its simplicity and satisfied efficiency.

When implementing our algorithm, we have heavily taking advantage of the library Thrust. An efficient counterpart of Thrust, CUB [23], has been developed recently. It was reported that CUB is faster than Thrust. We expect to gain a significant increase in overall performance of our algorithm by replacing Thrust with CUB. Future work should therefore include the implementation our algorithm using CUB and the evaluation of efficiency performance.

### Availability of CudaChain

CudaChain was implemented in C++ and the NVIDIA CUDA library and requires CUDA 4.0 or higher. The source code of CudaChain is available by sending a request to the first author via email.